\documentclass[useAMS,fleqn, usenatbib, final]{mnras}
\usepackage{lipsum}
\usepackage{multirow}
\usepackage[inline]{enumitem}
\usepackage{import}
\usepackage{tabularx}
\usepackage{glossaries}
\usepackage[titletoc,title]{appendix}
\usepackage{graphicx}
\usepackage{amsmath}
\usepackage{amssymb}
\usepackage{amsfonts}
\usepackage{mathtools}
\usepackage{bm}
\usepackage[utf8]{inputenc}
\usepackage[table]{xcolor}
\usepackage{url}
\usepackage[normalem]{ulem}

\newcommand{\num}{1}
\newcommand{\portsmouth}{Institute of Cosmology and Gravitation, University of Portsmouth, Portsmouth, PO1 3FX, UK}

\title[Consistent use of PHM models in GW analyses]{Reconsidering the consistent use of precessing, higher order multipole models for gravitational wave analyses}

\author[Hoy]{
\parbox{\textwidth}{
C.~Hoy$^{\num}$\thanks{charlie.hoy@port.ac.uk}
}
\vspace{0.2cm}\\
$^1$\portsmouth\\
\vspace{-1.5em}
}
\date{Accepted 2026 May 8. Received 2026 May 1; in original form 2026 February 4}
\pubyear{2025}

\begin{document}
\label{firstpage}
\pagerange{\pageref{firstpage}--\pageref{lastpage}}

\maketitle

\begin{abstract}
The growing number of gravitational-wave (GW) observations allows for constraints to be placed on the underlying population of black holes; current estimates show that black hole spins are small, with binaries more likely to have comparable component masses. Since general relativistic effects, such as spin-induced orbital precession and higher order multipole moments, are more likely to be observed for asymmetric binary systems, a direct measurement remains unlikely. Nevertheless, we continue to consistently probe these effects by performing Bayesian inference with our most accurate and computationally expensive models. As the number of GW detections increases, it may soon become infeasible to consistently use these models for analyses. In this paper, we provide a selection criterion that determines when less accurate and computationally cheaper models can be used without giving biased estimates for the population properties of black holes in the Universe. We show that when using our selection criterion, comparable estimates can be obtained for the underlying mass and spin distribution of black holes for a simulated ``worst-case'' scenario population, while reducing the overall cost of performing Bayesian inference on our population by $\sim 20\%$. We anticipate a reduction of up to $78\%$ in the overall cost for an astrophysically motivated population, since there are fewer events with observable spin-precession and higher order multipole power. 
\\
\end{abstract}


\section{Introduction}
The Advanced LIGO~\citep{LIGOScientific:2014pky}, Virgo~\citep{VIRGO:2014yos} and KAGRA~\citep{KAGRA:2020tym} gravitational-wave (GW) detectors have observed more than 200 signals from compact binary coalescences~\citep{LIGOScientific:2018mvr,LIGOScientific:2020ibl,LIGOScientific:2021usb,KAGRA:2021vkt,Nitz:2021zwj,Olsen:2022pin,Mehta:2023zlk,Wadekar:2023gea,LIGOScientific:2025slb}. Through a combination of Bayesian methods and highly accurate GW models~\citep[see e.g.][]{LIGOScientific:2025yae}, estimates for the individual source properties can be obtained, and combined to understand the underlying population properties of black holes in the Universe~\citep{LIGOScientific:2025pvj}. 

By combining the information from 158 GW signals, \cite{LIGOScientific:2025pvj} demonstrated that black holes have non-extremal spins ($\chi \sim 0.2$) that are preferentially aligned with the binaries orbit. They further highlighted that binary black holes are more likely to have equal component masses. Combined, this means that general relativistic effects -- such as spin-induced orbital precession and higher order multipoles -- are unlikely to be observed in GW signals. Spin-induced orbital precession occurs when the binary's spin is misaligned with the orbital angular momentum~\citep{Apostolatos:1994mx}, leaving characteristic amplitude and phase modulations in the emitted GW signal. Spin-precession is more likely to be observed in binaries with asymmetric component masses and large spin magnitudes~\citep{Fairhurst:2019vut,Fairhurst:2019srr}. Higher order multipole moments~\citep{Goldberg:1966uu,Thorne:1980ru} describe the amount of GW radiation emitted in a particular direction. The majority of power is emitted in the $(\ell, m) = (2, 2)$ quadrupole, but for asymmetric component masses higher order terms become observable in the GW signal~\citep{Mills:2020thr}. \cite{Hoy:2024wkc} demonstrated that up to the third GW observing run, only one event showed substantial
evidence for precession: GW200129\_065458~\citep{KAGRA:2021vkt,Hannam:2021pit}, and two events show substantial evidence for higher order multipoles: GW190412~\citep{LIGOScientific:2020stg} and GW190814~\citep{LIGOScientific:2020zkf}. Since then, other events may have shown signs of precession and higher order multipoles~\citep{LIGOScientific:2025slb}.

Although there is a low probability of observing general relativistic effects in a given signal~\citep[1 in 50 for spin-precession and 1 in 70 for higher order multipoles, see][]{Hoy:2024wkc}, the majority of GW observations to date\footnote{For GW candidates that likely originate from binary black hole mergers, it is standard to use models that incorporate spin-precession and higher order multipole moments. For neutron star black hole or binary neutron star mergers, models that incorporate tidal effects are used. Often these do not include spin-precession or higher order multipoles contributions~\citep[see e.g.][]{Dietrich:2019kaq,Thompson:2020nei,Matas:2020wab}.} are consistently analysed with models that incorporate spin-precession and higher order multipole moments~\citep{Nitz:2021zwj,LIGOScientific:2021usb,KAGRA:2021vkt,Olsen:2022pin,Mehta:2023zlk,LIGOScientific:2025slb}. This is natural since they more accurately describe numerical solutions to general relativity~\citep[see e.g.][]{MacUilliam:2024oif,Varma:2019csw,Estelles:2025zah,Hamilton:2025xru} and, as a result, they are less likely to produce biased source estimates~\citep[see e.g.][]{CalderonBustillo:2015lrt,Ramos-Buades:2023ehm,Thompson:2023ase,Yelikar:2024wzm,MacUilliam:2024oif,Dhani:2024jja,Akcay:2025rve}. However, when there is no measurable spin-precession and higher order multipole power in the observed signal, comparable estimates for the individual and collective properties of black holes are generally expected regardless of the physics included in the model. Indeed, \cite{Hoy:2025ule} demonstrated that the inferred expansion rate of the universe remains agnostic to the physics included in the waveform model, and \cite{Singh:2023aqh} demonstrated that neglecting higher order multipoles does not bias population inference. Given that simpler and less accurate models can reduce the computational cost of Bayesian analyses by six times~\citep[on average,][]{Hoy:2025ule}, a trivial method to reduce the growing computational cost of GW data analyses is to strategically use computationally cheaper models where possible.

In this work, we demonstrate that precessing, higher order multipole models do not need to be consistently used when analysing GW signals; rather, a combination of models can be used and comparable estimates for the underlying mass and spin distribution of black holes can be obtained. We propose a simple selection criterion that only uses computationally expensive and highly accurate models when known general relativistic effects are measurable in the observed signal; otherwise, computationally cheaper models that exclude spin-precession and/or higher order multipoles can be employed. We verify our selection criterion for a population of rapidly spinning black holes with a preference for asymmetric component masses; this non-astrophysically motivated population acts as a worse-case scenario where spin-precession and higher order multipole power is likely to be observed in a given GW signal. Based on our population, we estimate that our selection criterion reduces the overall cost of performing GW data analysis by at least $\sim 20\%$ without employing likelihood accelerated techniques~\citep{Cornish:2010kf,Canizares:2013ywa,Canizares:2014fya,Smith:2016qas,Vinciguerra:2017ngf,Zackay:2018qdy,Qi:2020lfr,Cornish:2021lje,Morisaki:2021ngj,Morisaki:2023kuq,Krishna:2023bug}. Given that $> 90\%$ of observed binary black holes have lower spin magnitudes and more symmetric component masses, an astrophysically motivated population will obtain larger efficiencies, and can likely use a stricter selection criterion without incurring biases.

It is possible that employing our selection criterion could reduce the computational cost further. For example, current catalogs produced by the LVK include parameter estimates from multiple waveform families; typically two or three~\citep{LIGOScientific:2025slb}. This is designed to mitigate the effect of waveform systematics, see~\cite{LIGOScientific:2025yae} for details (although see~\cite{Hoy:2024vpc} for an alternative technique). However, when precession and/or higher order multipole moments are not included, models from different waveform families generally agree well, see e.g.~\cite{Bohe:2016gbl}. This implies that repeating analyses with multiple families may not be necessary in all regions of the parameter space. Where our selection criterion indicates that precession and higher order multipole models are not required to reliably infer the parameters, potentially only one waveform family is required. As such, in reality, we may expect even larger reductions in computational cost.

This paper is structured as follows: in Sec.~\ref{sec:observable_gr_effects} we provide an overview of spin-induced orbital precession and higher order multipoles, and discuss the impact of neglecting these phenomena in models and parameter estimation. In Sec.~\ref{sec:selection_criteria} we describe our selection criterion and verify it in Sec.~\ref{sec:verification}. We finally conclude in Sec.~\ref{sec:discussion}.

\section{Observable general relativistic effects} \label{sec:observable_gr_effects}

\subsection{Spin-induced orbital precession}

In general relativity, a compact binary coalescence on a quasi-circular orbit is described by fifteen parameters when neglecting matter effects: eight describing the intrinsic properties of the source -- the compact masses, $m_{1}$ and $m_{2}$, the spin vectors of each object, $\mathbf{S}_{1}$ and $\mathbf{S}_{2}$ -- and seven describing the extrinsic properties -- the luminosity distance, inclination angle, right ascension, declination, polarization, phase and coalescence time. The total angular momentum of the binary $\mathbf{J}$ is defined as the sum of the orbital angular momentum $\mathbf{L}$ and the total spin of binary $\mathbf{S} = \mathbf{S}_{1} + \mathbf{S}_{2}$. If matter effects are included, as must be the case for neutron stars, additional parameters are required to describe the tidal deformability.

When $\mathbf{S}$ is misaligned with $\mathbf{L}$, the binary undergoes the general relativistic phenomenon of spin-induced orbital precession. In most cases, $\mathbf{L}$ precesses around the approximately constant $\mathbf{J}$ with an opening angle $\beta$. The strength of precession is described by the magnitude of the opening angle, with more prominent effects occurring for larger angles~\citep{Apostolatos:1994mx}. $\beta$ is primary determined by the total spin perpendicular to $\mathbf{L}$. For this reason, it is often convenient to define the effective perpendicular spin as,

\begin{equation}
    \chi_{\mathrm{p}} = \frac{1}{A_{1}}\max\left(A_{1} \chi_{1}\,\sin\theta_{1}, A_{2}\chi_{1}\,\sin\theta_{1}\right),
\end{equation}
where $A_{1} = 2 + 3m_{2}/(2m_{1})$, $A_{2} = 2 + 3m_{1}/2m_{2}$, $\chi_{i} = m_{i}^{2}|\mathbf{S}_{i}|$, and $\cos(\theta_{i}) = \mathbf{S}_{i} \cdot \mathbf{L}$~\citep{Schmidt:2014iyl}. $\chi_{\mathrm{p}}$ is defined between $0$ and $1$, where $0$ indicates a binary without spin-precession and $1$ indicates maximal spin-precession. Other effective perpendicular spins have also been proposed in the past~\citep{Thomas:2020uqj,Gerosa:2020aiw,Hoy:2024vpc}. It is also useful to define the effective parallel spin as~\citep{Racine:2008qv,Ajith:2009bn},

\begin{equation}
    \chi_{\mathrm{eff}} = \frac{m_{1}\chi_{1}\cos\theta_{1} + m_{2}\chi_{2}\cos\theta_{2}}{m_{1} + m_{2}}.
\end{equation}

\begin{figure*}
    \centering
    \includegraphics[width=0.95\textwidth]{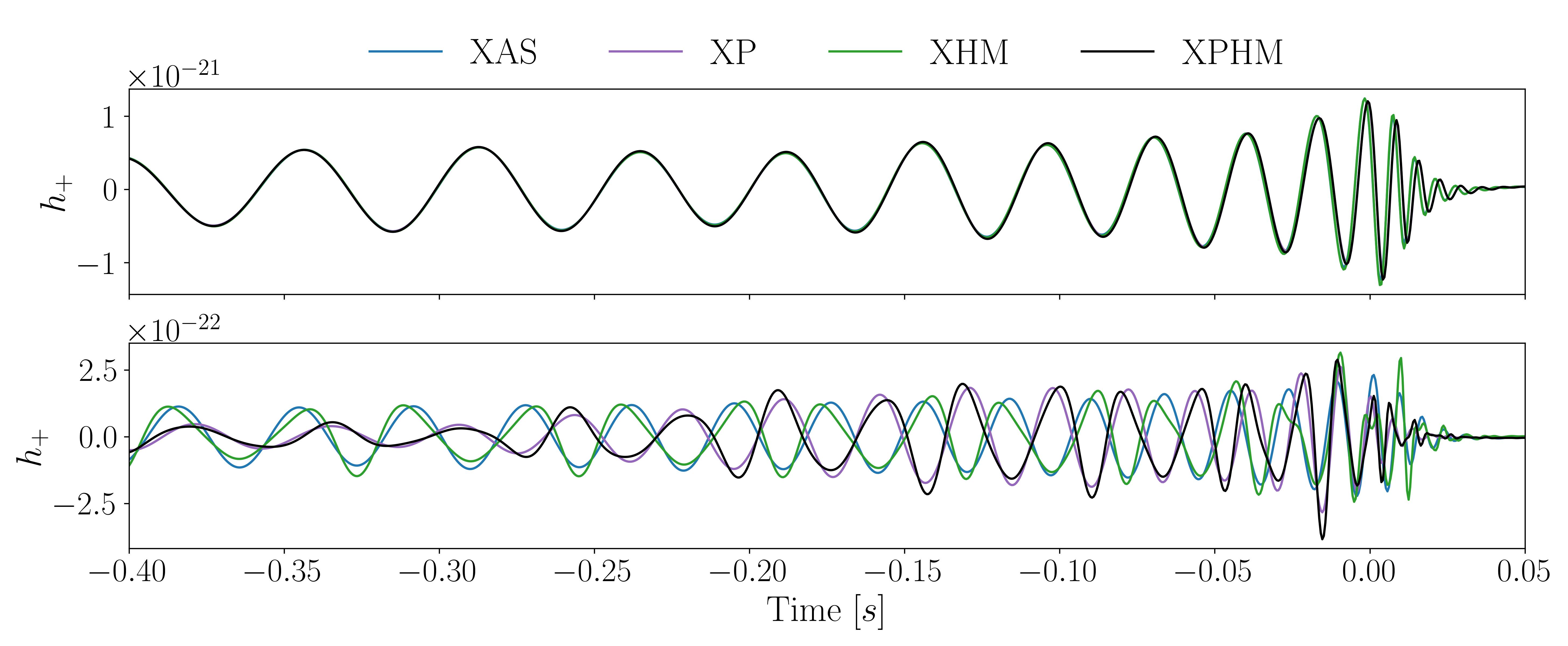}
    \caption{Plot showing the amplitude of the plus polarization, $h_{+}$, for two GW signals when assuming different physics; XAS assumes no spin-precession and higher order multipoles, XP assumes no higher order multipoles, XHM assumes no spin-precession, and XPHM includes both spin-precession and higher order multipoles. The \emph{Top} panel compares the GW signals for an equal mass ratio binary with small spin magnitudes, $\chi_{1} = \chi_{2} = 0.2$, observed at an inclination angle of $\pi / 6\, \mathrm{rad}$. This represents a binary that is consistent with the observed population, and therefore likely to be detected by the LVK~\citep{LIGOScientific:2025pvj}. The \emph{Bottom} panel compares the GW signals for a mass ratio $q = 0.1$ binary with large spin magnitudes, $\chi_{1} = \chi_{2} = 0.8$, observed at an inclination angle of $\pi / 2\, \mathrm{rad}$.}
    \label{fig:motivation_models}
\end{figure*}

While $\beta$ and $\chi_{\mathrm{p}}$ describe the intrinsic strength of precession, they are not sufficient to indicate whether it is observable given our GW detectors. This is because the characteristic amplitude and phase modulations associated with spin-induced orbital precession depend on the extrinsic properties of the source. The observability of spin-precession is described by the precession signal-to-noise ratio (SNR), $\rho_{p}$~\citep{Fairhurst:2019vut,Fairhurst:2019srr}. In the absence of precession, $\rho_{p}$ is expected to follow a $\chi$ distribution with two degrees of freedom from Gaussian noise alone. As such, in 90\% of GW signals with no observable precession, $\rho_{\mathrm{p}} < 2.1$. This null distribution is conservative and, in some cases, may prefer lower SNRs \citep[if informative priors are used, see][]{Hoy:2021dqg,Hoy:2024wkc}.

\subsection{Higher order multipole moments}

The observed GW signal $h = h_{+} + ih_{\times}$ can be decomposed into an infinite sum of multipoles using $-2$ spin-weighted spherical harmonics, ${}^{-2}Y_{\ell, m}$~\citep{Goldberg:1966uu, Thorne:1980ru}. Although the amplitude of each multipole varies across the parameter space, the $(\ell, m) = (2, 2)$ quadrupole typically dominates the total power of the signal~\citep{Mills:2020thr}, albeit not always~\citep{Fairhurst:2023beb,Ursell:2025ufb}. The amplitude of each multipole depends on the total mass and mass ratio $q = m_{2} / m_{1}$ of the binary, with the amplitude typically increasing for more asymmetric ($q < 1$), higher total mass binaries~\citep{Mishra:2016whh}. For most cases, the $(3,3)$ multipole is the most significant higher order contribution after the quadrupole. Similar to spin-precession, the observability of each multipole can be described by an SNR, $\rho_{\mathrm{\ell m}}$~\citep{Mills:2020thr}. We define $\rho_{\mathrm{HM}}$ as the largest multipole SNR beyond the quadrupole. $\rho_{\mathrm{HM}}$ is expected to follow the same null distribution as $\rho_{\mathrm{p}}$.

\subsection{Impact on models and parameter estimation} \label{sec:models}

\begin{figure*}
    \centering
    \includegraphics[width=0.93\textwidth]{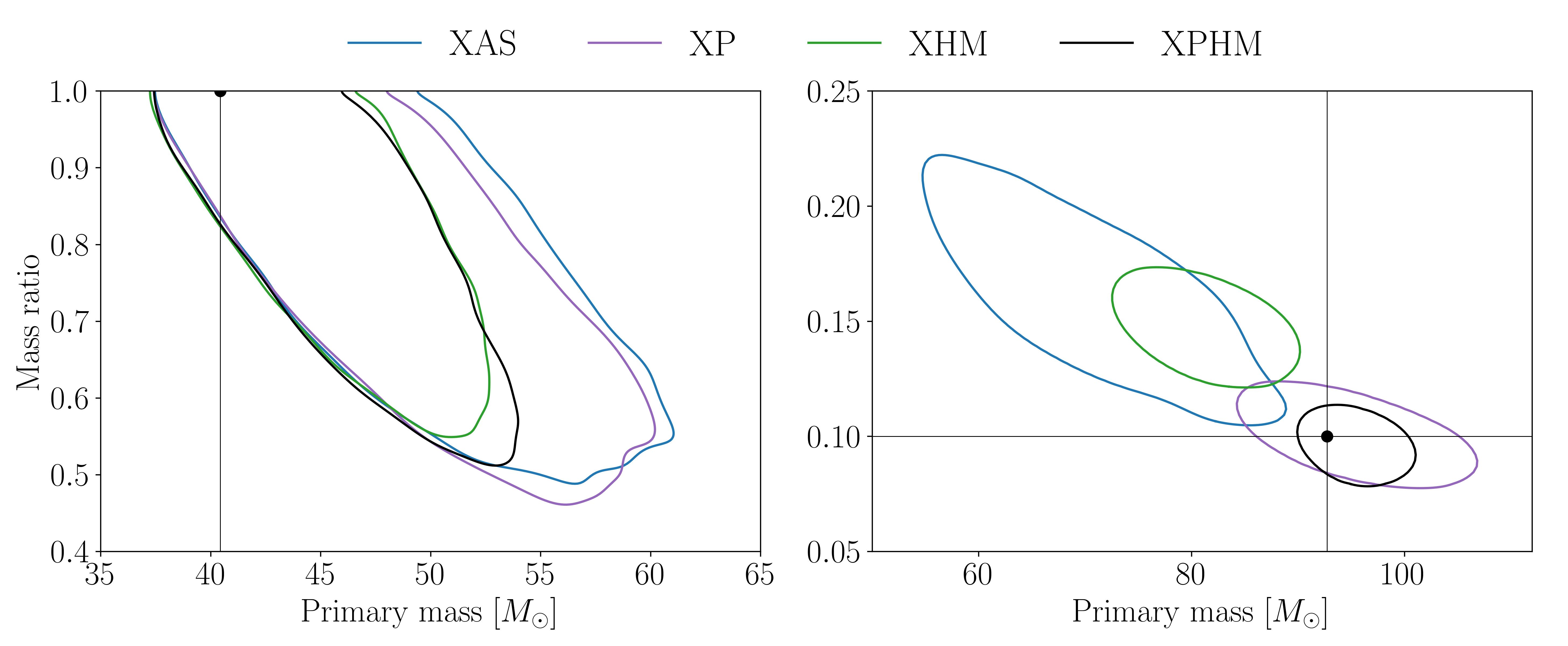}
    \caption{Plot showing the inferred primary mass and mass ratio when performing Bayesian inference on the GW signals shown in Fig.~\ref{fig:motivation_models}. We inject each GW signal into LIGO-Hanford and LIGO-Livingston, operating at their design densitivities~\citep{LIGOScientific:2014pky}. The \emph{Left} panel corresponds to a GW signal with no observable precession and higher order multipoles, the \emph{Top} panel in Fig.~\ref{fig:motivation_models}. The \emph{Right} panel corresponds to a GW signal with observable precession and higher order multipoles, the \emph{Bottom} panel in Fig.~\ref{fig:motivation_models}. In both cases we inject a simulated GW signal produced with XPHM into idealised Gaussian noise, and perform inference with XAS, XP, XHM and XPHM. The black cross hairs show the true values. The contours show the inferred 90\% credible interval.}
    \label{fig:single_event_pe}
\end{figure*}

GW models are typically constructed from a combination of analytic and semi-analytic approximations, as well as fits to numerical relativity calculations~\citep[see e.g.][]{LIGOScientific:2025yae}. Since model development is a labour intensive process, models that neglect spin-precession and multipoles beyond the dominant quadrupole are often developed first. In subsequent years, additional physics is added to more reliably describe numerical solutions to general relativity at the expense of computational efficiency. Numerous model families are available, with the LIGO--Virgo--KAGRA collaboration (LVK) typically using the {\texttt{Phenom}}~\citep{Husa:2015iqa, Khan:2015jqa, London:2017bcn, Hannam:2013oca, Khan:2018fmp, Khan:2019kot,Pratten:2020fqn, Garcia-Quiros:2020qpx, Pratten:2020ceb,Estelles:2020osj, Estelles:2020twz, Estelles:2021gvs,Hamilton:2021pkf,Thompson:2023ase,Colleoni:2024knd,Hamilton:2025xru}, {\texttt{SEOBNR}}~\citep{Bohe:2016gbl, Cotesta:2018fcv, Cotesta:2020qhw, Ossokine:2020kjp,Babak:2016tgq,Pan:2013rra,Pompili:2023tna,Ramos-Buades:2023ehm,Gamboa:2024hli} and {\texttt{NRSurrogate}}~\citep{Varma:2018mmi,Varma:2019csw} families. Focusing on the ``X'' generation of {\texttt{Phenom}} waveforms, a GW signal from a quasi-circular binary black hole can be described by
\begin{enumerate*}
    \item {\texttt{PhenomXAS}} (XAS) when neglecting spin precession and higher order multipoles~\citep{Pratten:2020fqn},
    \item {\texttt{PhenomXP}} (XP) when neglecting higher order multipoles~\citep{Pratten:2020ceb},
    \item {\texttt{PhenomXHM}} (XHM) when neglecting spin-precession~\citep{Garcia-Quiros:2020qpx} and,
    \item {\texttt{PhenomXPHM}} (XPHM) otherwise~\citep{Pratten:2020ceb}. 
\end{enumerate*}
On average\footnote{We averaged over 5000 waveform evaluations with chirp masses in the range $2 < \mathcal{M} < 200$ and mass ratios in the range $0.05 < q < 1$. We chose the duration of each waveform evaluation based on the component masses and spins. We consistently used a maximum frequency of $1024\,\mathrm{Hz}$ and a sampling frequency of $2048\,\mathrm{Hz}$.} XPHM is $1.2\times$ slower than XAS, $1.1\times$ slower than XP, and $1.2\times$ slower than XHM for a single waveform evaluation.

Since both the intrinsic and extrinsic properties of the source affect the observability of spin-precession and higher order multipoles in the detected signal, the intrinsic $\rho_{\mathrm{p}}$ and $\rho_{\mathrm{HM}}$ can be approximately zero for certain configurations. For these cases, a GW signal produced with XPHM is indistinguishable from XAS~\citep[see e.g.][for details on indistinguishability in the GW context, and it's dependence on the SNR of the signal]{Thompson:2025hhc}; in other words, there is little advantage in using more computationally expensive models to describe the observed GW signal. This is shown in the top panel of Fig.~\ref{fig:motivation_models}. We see that for a binary typically observed by the LVK~\citep{LIGOScientific:2025pvj} -- an equal mass ratio binary with small spin magnitudes ($\chi_{1} = \chi_{2} = 0.2$), $\rho_{\mathrm{p}} = 0.5$, $\rho_{\mathrm{HM}} = 0.5$, and observed at a total network SNR of 20 -- the predicted GW signals from different models are indistinguishable, despite missing key physical effects in most cases. Although demonstrated with the {\texttt{Phenom}} family, the same result would be obtained with {\texttt{SEOBNR}} and {\texttt{NRSurrogate}}. Of course, this is not always the case. For a binary with significant evidence for spin-precession and higher order multipoles, $\rho_{\mathrm{p}},\, \rho_{\mathrm{HM}} > 2.1$, we see significant differences in the GW signals, as shown in the bottom panel of Fig.~\ref{fig:motivation_models}. To observe this level of disagreement requires extreme binary configurations that are unlikely to be observed given current expectations for the mass and spin distribution of black holes in the Universe~\citep{LIGOScientific:2025pvj} -- here we show a mass ratio $q = 0.1$ binary with large spin magnitudes ($\chi_{1} = \chi_{2} = 0.8$), $\rho_{\mathrm{p}} = 7.9$, $\rho_{\mathrm{HM}} = 7.8$, and observed at a total network SNR of 20. 

As our ability to infer the \emph{true} source properties from an observed GW signal depends on the chosen model~\citep[see][and App.~\ref{sec:bayesian_inference} for details]{LIGOScientific:2025yae}, we expect to obtain comparable source estimates when analysing a GW signal with no evidence for spin-precession and higher order multipole moments regardless of the model used. We note that for some cases, the lack of evidence for precession and higher order multipoles can help further constrain the parameters. GW190814~\citep{LIGOScientific:2020zkf} clearly demonstrates this: the lack of evidence for precession reduces the uncertainty in the inferred mass of the secondary object from $2.4-2.8\, M_{\odot}$ to $2.5-2.7\, M_{\odot}$ (at the 90\% credible interval), see Fig.~4 in~\cite{LIGOScientific:2020zkf}. Although the uncertainties can shrink, the median is often comparable. This becomes more relevant for signals with large total SNR.

To demonstrate that we obtain comparable source estimates when there is no evidence for precession and higher order multipoles in the signal, we inject the XPHM signal shown in the top panel of Fig.~\ref{fig:motivation_models} into idealised Gaussian noise for a network of 2 advanced LIGO~\citep{LIGOScientific:2014pky} detectors. We then perform Bayesian inference with the {\texttt{dynesty}}~\citep{Speagle:2019ivv} nested sampler via {\texttt{bilby}}~\citep{Ashton:2018jfp, Romero-Shaw:2020owr} with XAS, XP, XHM and XPHM. In all cases, we ensure wide and agnostic priors for all parameters and 2000 live points to ensure robust results; further details can be found in App.~\ref{sec:bayesian_inference}. In the left panel of Fig.~\ref{fig:single_event_pe} we show the two-dimensional marginalized posterior distribution for $m_{1}$ and $q$. We see that we obtain comparable estimates for source properties regardless of the model used, as expected. This is despite XAS, XP and XHM using $5\times$, $2\times$ and $4\times$ less computational resources than XPHM respectively\footnote{Although a single waveform evaluation for \emph{e.g.} XAS is only $1.2\times$ faster than XPHM, $\mathcal{O}(10^{7})$ waveform evaluations are performed in a typical Bayesian analysis, hence the significant reduction in computational resources. We note that XAS also explores a reduced parameter space compared to XPHM (4 dimensions fewer), due to neglecting in-plane spin components. This also contributes to the reduction in computational cost but not as significantly as the reduction in waveform evaluation time. When sampling over the exact same space, XAS uses $3.5\times$ less computational resources than XPHM.} When performing model selection, see App.~\ref{sec:bayesian_inference} for details, we see that the $\log_{10}$ Bayes factor's between the different models are less than 1, highlighting no model preference. \cite{Hoy:2025ule} demonstrated that across a population of simulated signals, XAS uses on average $6\times$ less computational resources than XPHM. The exact computational cost varies for each signal as it depends on priors, data durations, frequencies considered etc.

For cases where we observe large differences between predicted signals produced with different models, \emph{i.e.} for binaries with significant evidence for spin-precession and higher order multipoles, we expect to observe biases in the source properties when employing models that neglect physics. This is shown in the right hand panel of Fig.~\ref{fig:single_event_pe}. We see that when injecting the XPHM signal shown in the bottom panel of Fig.~\ref{fig:motivation_models} into idealised Gaussian noise and recovering with XAS, XHM, XP and XPHM, XAS and XHM recover a biased primary mass and mass ratio. XP and XPHM recover the true values within the inferred 90\% credible interval, implying that neglecting higher order contributions to the total SNR of the signal is key to biasing our inference. This is well known and has been previously demonstrated~\citep[see e.g.][]{Chatziioannou:2014coa,Purrer:2015nkh,CalderonBustillo:2015lrt,London:2017bcn,Cotesta:2018fcv,Kalaghatgi:2019log,Pratten:2020igi,Garcia-Quiros:2020qpx, Krishnendu:2021cyi}. This is supported by $
\log_{10}$ Bayes factor's between the different models; XPHM is preferred over XAS with a $\log_{10}\mathcal{B} = 40$.

\section{Determining the selection criteria} \label{sec:selection_criteria}

Based on the results in Sec.~\ref{sec:observable_gr_effects}, a trivial method for reducing the computational cost of performing Bayesian inference for a population of GW signals (without employing any new techniques) is to design a selection criteria that strategically uses XAS when there is no evidence for precession and higher order multipoles in the detected signal, and XP (XHM) when there is significant evidence for precession (higher order multipoles). This means that we only use the most computationally expensive model, XPHM, for the few signals where necessary. By using only models that contain sufficient physics to describe the observed signal, biases can be controlled, and computational cost reduced. Based on Figs.~\ref{fig:motivation_models} and \ref{fig:single_event_pe}, a natural strategy is to set an SNR threshold based on $\rho_{\mathrm{p}}$ and $\rho_{\mathrm{HM}}$; if $\rho_{\mathrm{p}}$ and $\rho_{\mathrm{HM}}$ are small, XAS is sufficient to describe the signal, while if $\rho_{\mathrm{p}}$ and $\rho_{\mathrm{HM}}$ are large, XPHM is essential. Of course, this requires an estimate for $\rho_{\mathrm{p}}$ and $\rho_{\mathrm{HM}}$ \emph{a priori}.

\cite{Fairhurst:2023idl} demonstrated that {\texttt{simple-pe}} can directly extract $\rho_{\mathrm{p}}$ and $\rho_{\mathrm{HM}}$ from the GW strain data through matched filtering~\citep{Brown:2004vh}, within seconds on a single CPU. The process is as follows: first, the best fitting dominant quadrupole-only waveform that maximises the
matched filter SNR of the observed signal is identified through an optimisation step. A similar strategy has also been employed in low-latency searches~\citep{Nitz:2018rgo}. This \emph{template} is split into a series orthogonal components, with each describing the isolated contribution from precession~\citep{Fairhurst:2019vut} and higher order multipoles~\cite{Mills:2020thr,Fairhurst:2023idl}. Each component is then matched filtered against the data to obtain an SNR. A similar procedure has been implemented in {\texttt{PyCBC}}~\citep{Usman:2015kfa} to search for precessing binaries~\citep{McIsaac:2023ijd, Harry:2025eqk}.

It is important to choose a selection criteria that \emph{e.g.} ensures XAS is only used for analyses when there is no evidence for precession and higher order multipoles in the signal. Otherwise, biases are expected, see Fig.~\ref{fig:single_event_pe}. Therefore, we first compare the matched filtered $\rho_{\mathrm{p}}$ and $\rho_{\mathrm{HM}}$ against the true SNRs to identify whether they can reliably be used to define a threshold. We are only interested in false negatives: cases where we identify the signal with no evidence for precession and/or higher order multipoles despite the true signal showing strong observable features. False negatives imply that we may select a model that is insufficient to describe the signal. False positives are tolerable since this implies that a model with more physics than necessary will be used.

\begin{figure}
    \centering
    \includegraphics[width=0.45\textwidth]{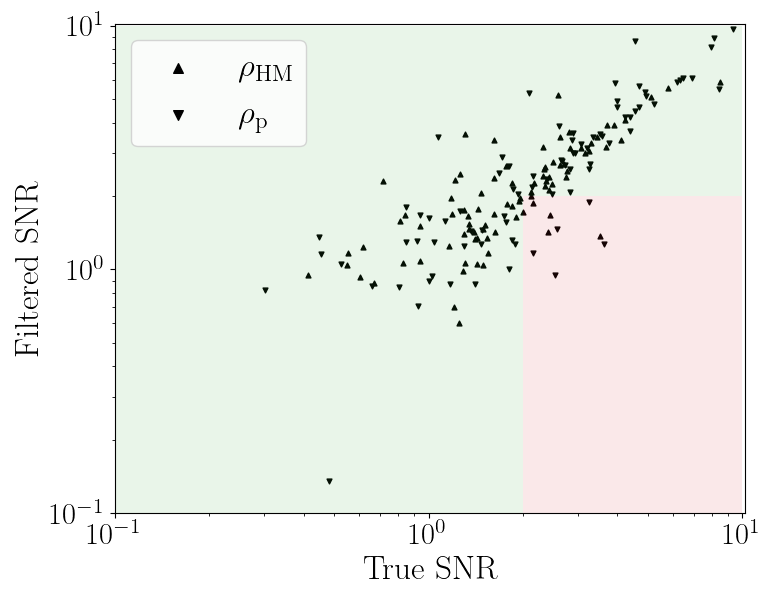}
    \caption{Plot comparing the matched filtered SNR in precession $\rho_{\mathrm{p}}$ and higher order multipoles $\rho_{\mathrm{HM}}$ to the true SNR for our worst-case scenario population. The shaded red region shows false negatives: a region where we identify the binary as having no evidence for precession and/or higher order multipoles from the filtered SNR despite there being observable features in the signal. The green region shows false positives, true positives and true negatives. We define $\rho_{\mathrm{p}} = \rho_{\mathrm{HM}} = 2.0$ for a binary with significant evidence of precession and/or higher order multipoles. This defines the boundary for the different shaded regions in the figure.}
    \label{fig:snr_accuracy}
\end{figure}

We simulate a \emph{worst-case scenario} population of binary black holes: we consider a population of 90 binary black holes with large spin magnitudes and isotropic spin tilts -- leading to significant evidence for precession -- with a preference for asymmetric masses -- leading to significant evidence for higher order multipoles. The LVK has started to see evidence that these binaries may exist in nature~\citep{LIGOScientific:2025brd}. We only draw binaries with observable GW signals given current detector networks. Rather than injecting signals into real GW data and performing a search to identify which are detected~\citep[see e.g.][and App.~\ref{sec:pop_inference} for details]{LIGOScientific:2025pvj}, we simplify the process by assuming signals with total SNR greater than 12 are observable~\citep{Essick:2023toz,Mould:2023eca,Gerosa:2024isl,Agarwal:2024hld,Hoy:2025ule}. We calculate that $\sim 8\%$ of binaries drawn from the observed LVK population~\citep{LIGOScientific:2025pvj} are included in our simulated population. We then generate GW signals for each binary merger with XPHM, and inject all into different realisations of idealised Gaussian noise for an Advanced LIGO and Advanced Virgo network operating at $75\%$ duty cycle. We note that the LVK have observed $\sim 60$ signals with total (median) SNR greater than 12. By considering a population of 90 simulated signals with total SNR greater than 12 in this work, we therefore remain consistent with current, and near future detector sensitivities~\citep{LIGOScientific:2025slb}. We consider this worst-case scenario population since it increases the number of GW signals with observable $\rho_{\mathrm{p}}$ and $\rho_{\mathrm{HM}}$ compared to current population expectations.

We obtain our simulated population by assuming that the mass ratio follows a power law $q^{\beta_{q}}$ with index $\beta_{q} = -1.1$, and spin magnitudes follow a truncated Gaussian distribution with mean $a_{i} = 0.7$ and standard deviation $\sigma=0.2$. We also assume that the primary mass follows a truncated power law with index $\alpha=-2.3$, between a minimum, $m_{\mathrm{min}} = 5\, M_{\odot}$,and maximum $m_{\mathrm{max}} = 80\, M_{\odot}$ mass. We assume that there is a Gaussian component in the mass distribution at $m_{1} = 34\, M_{\odot}$, with relative weight $\lambda = 0.038$ and standard deviation $\sigma_{\rm g} = 5~M_{\odot}$. This is the same simulated population used in~\cite{Hoy:2025ule} and inspired from previous LVK results~\citep{KAGRA:2021duu}.

As shown in Fig.~\ref{fig:snr_accuracy}, {\texttt{simple-pe}} approximately recovers the expected linear relationship between the true and matched filter SNRs, especially at large values ($\mathrm{SNR} \gtrsim 2)$. This implies that we are unlikely to misidentify the signal when there is extreme evidence for spin-precession and higher order multipoles in the signal. At low SNRs, the linear relationship starts to break down, which is expected due to the presence of Gaussian noise. We see that if $\rho_{\mathrm{p}} = \rho_{\mathrm{HM}} = 2.0$ corresponds to a binary with significant evidence of precession and/or higher order multipoles, only $6\%$ of matched filter SNRs lead to false negatives, with most lying near the boundary. When repeating this analysis with the same simulations injected into \emph{zero noise}\footnote{The noise is set to its expectation value under infinite Gaussian
noise realizations}, we find that the number of false negatives decreases to 3\%.

There are several false negatives where \emph{e.g.} the true SNR is $\sim 3.5$ while we obtain a matched filter SNR $\sim 1.5$. One of these has significant evidence spin-precession. Since we misidentify the significance of this signal, we specifically show the impact of the chosen model on the posterior for this event in App.~\ref{sec:misidentified_signal}. We show that using a computationally cheaper model still recovers the true values within the 90\% credible interval. This gives confidence that even if the signal is misidentified, we obtain reasonable parameter estimates.

Given that we consider a worst-case scenario population, we expect this to be an upper limit. Since the main driving force behind the false negatives seems to be a consequence of the specific Gaussian noise realisation chosen, we conclude that the matched filter $\rho_{\mathrm{p}}$ and $\rho_{\mathrm{HM}}$ are robust to define a SNR threshold.

\section{Identifying an SNR threshold} \label{sec:verification}

Our aim is to define a selection criteria -- and hence an SNR threshold for $\rho_{\mathrm{p}}$ and $\rho_{\mathrm{HM}}$ -- such that when hierarchical Bayesian inference~\citep[see][ and App.~\ref{sec:pop_inference} for details]{Thrane:2018qnx} is performed on a population of GW signals, the inferred mass and spin distribution of black holes remains unchanged. By doing so, this implies that our ability to extract key physics from our GW observations remains unaltered. 

To identify a suitable threshold, $\rho_{\mathrm{thres}}$, we analyse all 90 binary black holes in our simulated, worst-case scenario population described in Sec.~\ref{sec:selection_criteria} numerous times: first when consistently using XPHM for all single-event Bayesian analyses, and for the others: when using a model chosen from a threshold on the matched filtered $\rho_{\mathrm{p}}$ and $\rho_{\mathrm{HM}}$ from {\texttt{simple-pe}}. We chose 6 different thresholds, $\rho_{\mathrm{thres}} \in [0.5, 1.0, 1.5, 2.0, 2.5, 3.0]$. We use,

\begin{enumerate}[leftmargin=2em]
    \item XAS when $\rho_{\mathrm{p}} < \rho_{\mathrm{thres}}$ and $\rho_{\mathrm{HM}} < \rho_{\mathrm{thres}}$,
    \item XHM when $\rho_{\mathrm{p}} < \rho_{\mathrm{thres}}$ and $\rho_{\mathrm{HM}} > \rho_{\mathrm{thres}}$,
    \item XP when $\rho_{\mathrm{p}} > \rho_{\mathrm{thres}}$ and $\rho_{\mathrm{HM}} < \rho_{\mathrm{thres}}$,
    \item XPHM when $\rho_{\mathrm{p}} > \rho_{\mathrm{thres}}$ and $\rho_{\mathrm{HM}} > \rho_{\mathrm{thres}}$.
\end{enumerate}
Using the posterior distributions obtained for each event, we perform hierarchical inference with the {\texttt{dynesty}} nested sampler via {\texttt{gwpopulation}}~\citep{2019PhRvD.100d3030T}, as is common~\citep{LIGOScientific:2025pvj}. We consistently assume a {\textsc{Powerlaw+Peak}} model~\citep{LIGOScientific:2020kqk} for the primary component mass, and a power law for the mass ratio distribution. We also assume that the component spin magnitudes are independently and identically drawn from a Beta distribution~\citep{Wysocki:2018mpo}, and a mixture model comprising two components for the spin tilt is used:
an isotropic component, and a second component
in which the spins are preferentially aligned with
$\mathbf{L}$~\citep{Talbot:2018cva}. In all cases, we ensure wide and agnostic priors for all parameters and 1000 live points. More details are provided in App.~\ref{sec:pop_inference}.

\begin{figure*}
    \centering
    \includegraphics[width=0.93\textwidth]{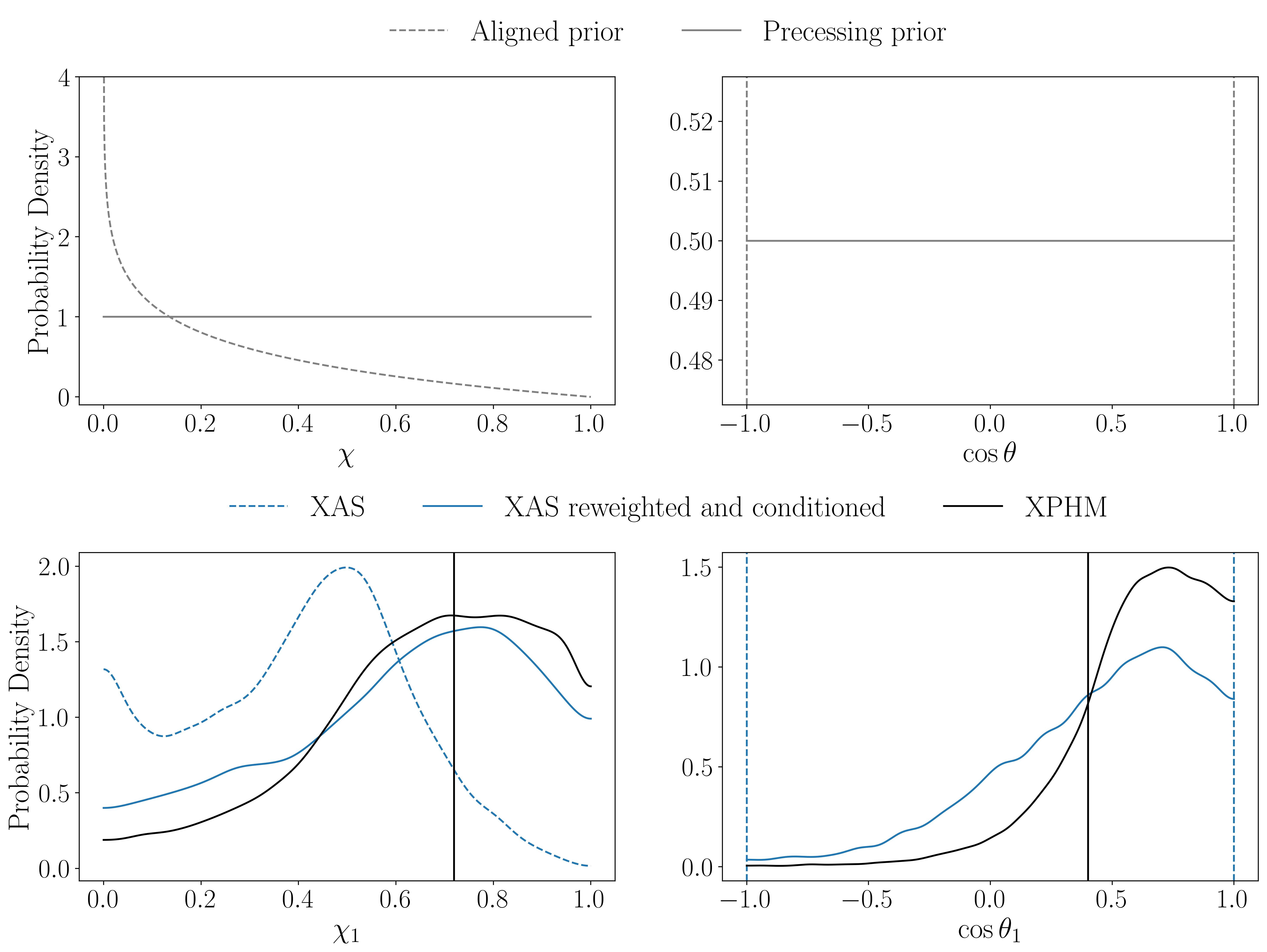}
    \caption{Plot showing the reweighting and conditioning procedure that is applied to the spin magnitude, $\chi$, and spin tilt, $\cos\theta$, posterior distributions obtained with XAS and XHM. The posterior distributions were obtained by analysing a simulated GW signal injected into idealised Gaussian noise. The binary had component masses $m_{1} = 34, M_{\odot}, m_{2} = 34\, M_{\odot}$, spin magnitudes $\chi_{1}=0.7, \chi_{2}=0.8$ and spin tilts $\cos\theta_{1} = 0.4, \cos\theta_{2} = 0.3$. The top row highlights the different spin-priors used for XAS/XHM (grey dash) and XP/XPHM (grey solid). The bottom row compares the original XAS (blue dashed) and the XAS reweighted and conditioned posterior distribution (blue solid) with XPHM (black solid). In the bottom row, the black vertical line indicates with the true value.}
    \label{fig:rweighting}
\end{figure*}

\begin{figure*}
    \centering
    \includegraphics[width=0.96\textwidth]{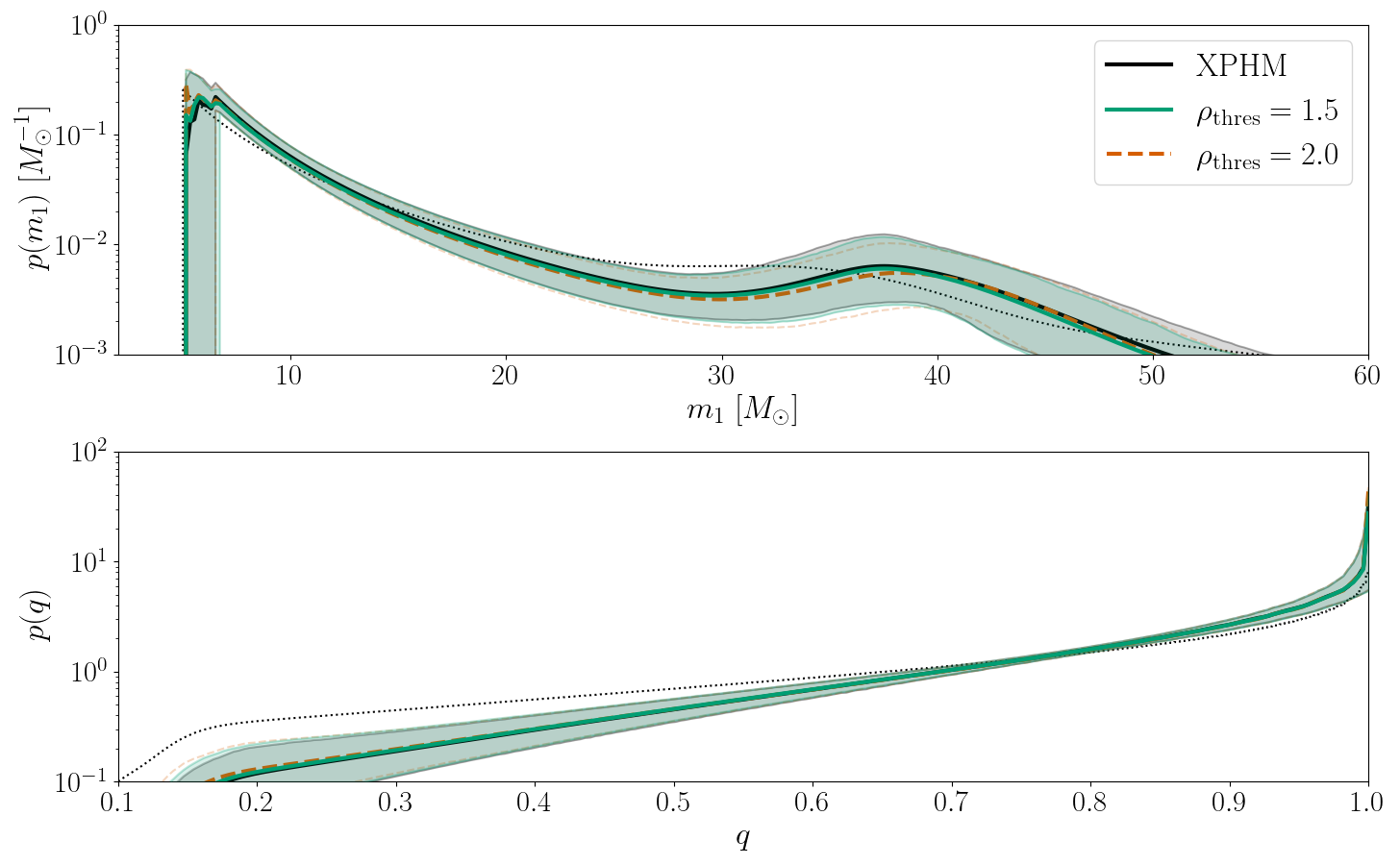}
    \includegraphics[width=0.96\textwidth]{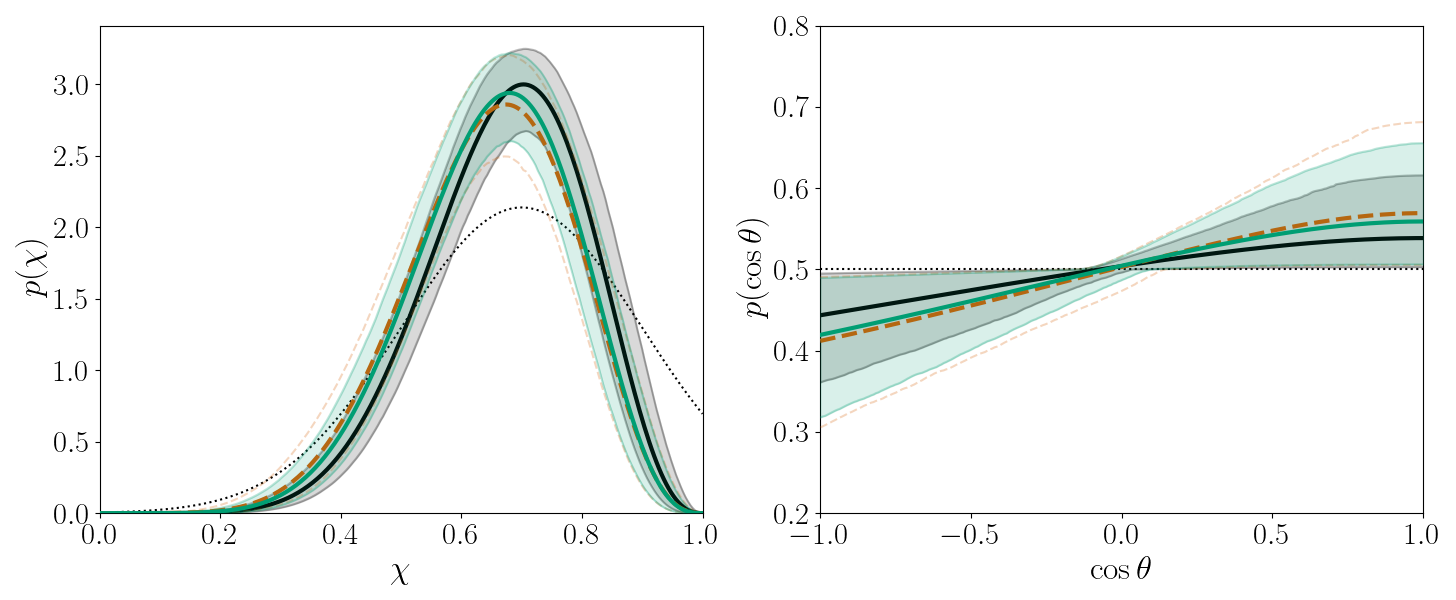}
    \caption{Plot comparing the inferred mass and spin distribution for a simulated highly spinning population. The top and middle panels show the inferred distribution for the primary mass $m_{1}$ and mass ratio $q = m_{2} / m_{1}$ respectively. The bottom left and bottom right panels show the inferred spin magnitude, $\chi$, and spin orientation, $\cos\theta$, respectively. The black solid line shows the distribution inferred when solely using posterior samples obtained with XPHM. The green solid line shows the distribution inferred when using an SNR threshold of $\rho_{\mathrm{thres}} = 1.5$, and the orange dashed line shows the distribution inferred when using a stricter SNR threshold of $\rho_{\mathrm{thres}} = 2.0$. The dotted line shows the true distribution of the simulated population, and the bands show the inferred 90\% credible intervals.}
    \label{fig:pop_inf}
\end{figure*}

An issue with combining single-event posterior distributions obtained with different models is that often alternative spin-priors are used; for models that neglect spin-precession (XAS and XHM) we typically use an aligned spin prior that is uniform in spin magnitude after marginalizing the in-plane spin contributions~\citep{Lange:2018pyp}, while models that include spin-precession (XP and XPHM) typically use a uniform in spin magnitude prior. This results in XAS/XHM typically inferring lower spin magnitudes and exactly aligned (0) or anti-aligned ($\pi$) spin-tilts, see Fig.~\ref{fig:rweighting}. If naively combined with the posterior distributions obtained with XP and XPHM, biases are expected in the inferred spin-distribution of black holes. Similarly, we expect biases in the mass distribution due to the well known mass-spin degeneracy~\citep{Baird:2012cu}. To overcome this, we \emph{predict} the in-plane spin distributions given the posteriors obtained with XAS and XHM.

In more detail, we re-weight the spin-magnitude distributions from the aligned-spin prior to a uniform in spin-magnitude prior. We then assume a uniform distribution for the spin-tilts (same as the one used for XP and XPHM) and condition them on the observed aligned-spin of the binary, $\chi_{\mathrm{eff}}$. In Fig.~\ref{fig:rweighting}, we show the result of this process. We see that after re-weighing and conditioning the XAS distributions, we obtain posteriors that are comparable to XPHM for signals with moderate observable precession and higher order multipoles. For this case, $\rho_{\mathrm{p}} = 2.0$ and $\rho_{\mathrm{HM}} = 0.6$. However, this process breaks down when there is significant spin-precession in the GW signal. This is expected since our technique is essentially ignoring any measurable information from precession and, as a result, the inferred spin magnitudes are smaller and spin tilts more aligned. We note that this is not an issue in our analysis, as we propose only using XAS and XHM for signals without observable $\rho_{\mathrm{p}}$ and $\rho_{\mathrm{HM}}$.

Our re-weighing and conditioning technique (applied to the XAS/XHM posteriors) is similar in principle to sampling over the priors used by XP and XPHM, and only considering the aligned-spin components during the inference. However, there will be small differences because our technique is sampling from a prior conditioned on the observed $\chi_{\mathrm{eff}}$, rather than remaining agnostic. In our testing, our reweighting and conditioning algorithm obtained results that were more comparable to XPHM for the primary spin magnitude. An alternative, more robust, method could be to use Sequential Monte Carlo techniques~\citep{Williams:2025aar}. However, our re-weighing and conditioning algorithm uses significantly fewer computational resources and is sufficient for our need; our reweighting and conditioning technique takes $\mathcal{O}(1)$ CPU minute to complete.

Fig.~\ref{fig:pop_inf} shows our key result: when using a threshold of $\rho_{\mathrm{thres}} = 1.5$, the inferred mass and spin distributions remain comparable to XPHM (within the 90\% credible intervals), and the astrophysical understanding of black holes remains the same. We see that both distributions recover the properties of the injected population well, although there are some small discrepancies caused by the limited number of GW signals considered; we see that the inferred mass ratio distribution slightly underestimates the number of highly asymmetric component mass binaries ($q \lesssim 0.15$), which is likely due to the random process of drawing 90 observable binary parameters from the simulated population. We also see that the spin peak is recovered well despite being modelled as a Beta distribution. However, it naturally goes to $0$ at maximal spin, unlike the true truncated Gaussian. See App B.5.1 in~\cite{LIGOScientific:2025pvj} for a discussion of the limitations of the Beta distribution. A truncated Gaussian could have been used rather than a Beta distribution for inference -- to match the true simulated population -- but we used a Beta distribution as it has been commonly used in the literature~\citep[see e.g.][]{LIGOScientific:2020kqk,KAGRA:2021duu}. It is therefore expected that our inference did not recover the true spin distribution. We note that we obtain marginally more support for aligned spin binaries compared to XPHM. 

The uncertainties on the inferred mass and spin distribution are expected to shrink if our simulated population contained $\mathcal{O}(1000)$ binaries~\citep{Mould:2025dts,Vitale:2025lms}. However, we note that a population of this size may exhibit biased themselves owing to uncontrolled Monte-Carlo errors when evaluating the likelihood, see~\cite{Essick:2022ojx,Talbot:2023pex,Heinzel:2025ogf} for details. As a result, even if a population of $\mathcal{O}(1000)$ GW signals was observed, we may be forced to analyse a smaller number of higher SNR signals to control this growing error~\citep{Wolfe:2025yxu}. Therefore, it is unlikely that the uncertainties will shrink enough to expose differences between XPHM and when imposing a selection threshold $\rho_{\mathrm{thres}} = 1.5$. In the event that we suspect they will, a smaller threshold can be used.

\begin{figure}
    \centering
    \includegraphics[width=0.45\textwidth]{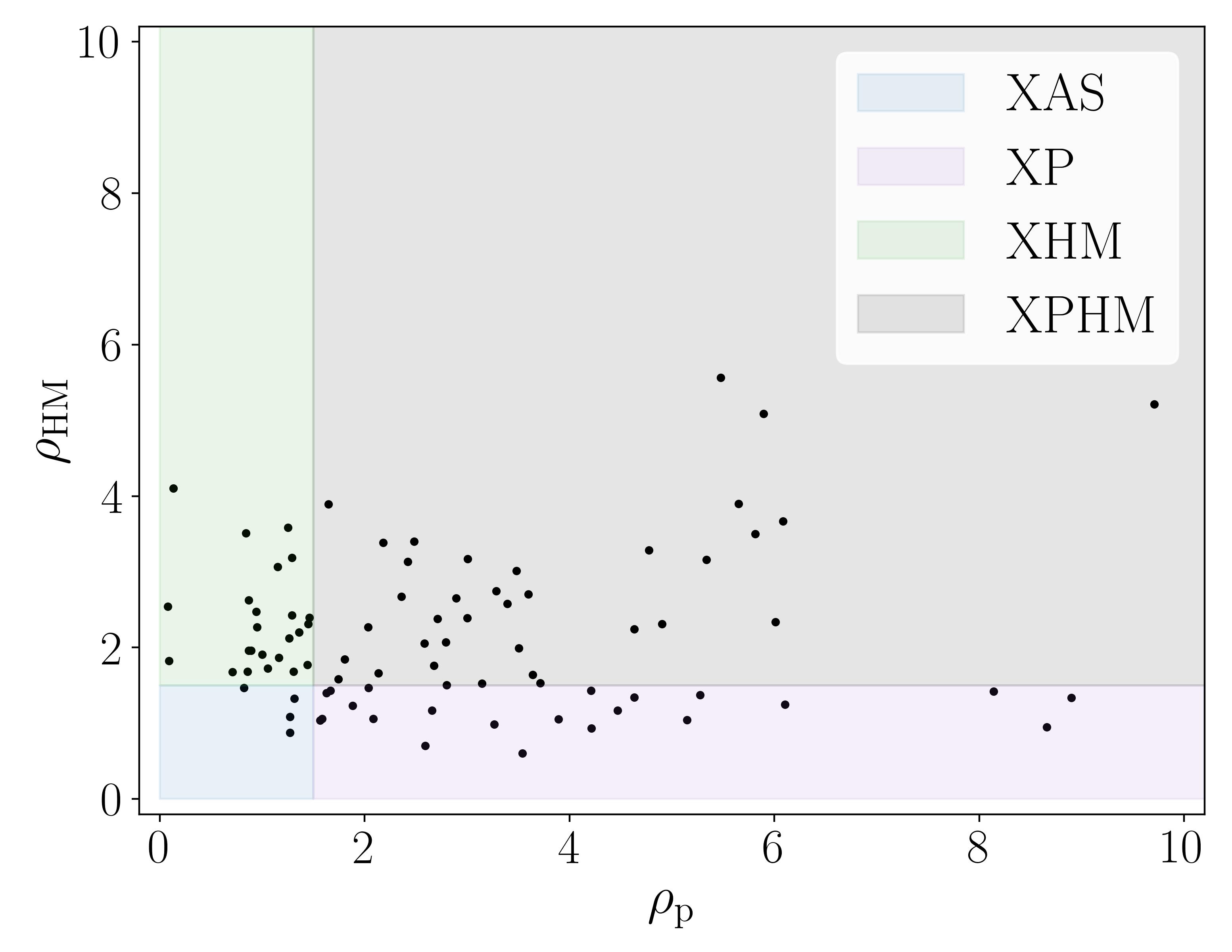}
    \caption{Plot showing the GW signals in our simulated, worst-case scenario population that were analysed by each model. We show this as a function of $\rho_{\mathrm{p}}$ and $\rho_{\mathrm{HM}}$. We use the recommended SNR threshold of $\rho_{\mathrm{thres}} = 1.5$.}
    \label{fig:SNR_dist}
\end{figure}

From Fig.~\ref{fig:SNR_dist} we see that when using $\rho_{\mathrm{thres}} = 1.5$, 4 events were analysed with XAS, 22 with XP, 24 with XHM and 38 with XPHM. We calculate that choosing the most sufficient and cheapest model to describe the observed signal reduces the overall cost of performing Bayesian inference on the population by $\sim 20\%$: a total of 40 CPU years compared to 50 CPU years when consistently using XPHM for all signals.

We see that when the SNR threshold is increased to $\rho_{\mathrm{thres}} = 2.0$, noticeable differences are seen in the inferred spin distribution, particularly in the 90\% credible interval of the spin tilt with marginally more support for aligned spin binary's. This is despite the mass distribution remaining comparable to XPHM. Given that a greater number of GW signals will be analysed with XAS/XHM for a higher threshold, this is consistent with the re-weighing and conditioning process underestimating the true spin-magnitude and spin tilt distributions. This is more clearly seen in Fig.~\ref{fig:pop_inf_multiple_snrs} in App.~\ref{sec:different_thresholds}. As the threshold is further increased, more significant biases in the spin distribution is seen, with a preference for lower spin magnitudes and more aligned binaries. For this reason, although a threshold of $\rho_{\mathrm{thres}} = 2.0$ reduces the computational cost by $\sim 25\%$, we recommend using a maximum threshold of $\rho_{\mathrm{thres}} = 1.5$ to mitigate the risk of biased results.

\section{Discussion} \label{sec:discussion}

In this work, we investigated a method for reducing the computational cost of performing Bayesian inference on a population of GW signals. Rather than consistently using highly accurate and computationally expensive models for GW analyses -- as is the standard in the LVK and other work -- we developed a selection criteria to determine when less accurate, and computationally cheaper models can be used without incurring biases in both single-event and hierarchical Bayesian analyses. We showed that by using an SNR threshold of $\rho_{\mathrm{thres}} = 1.5$ on the matched filtered $\rho_{\mathrm{p}}$ and $\rho_{\mathrm{HM}}$, the inferred mass and spin distributions for a worse-case scenario population of binary black holes remain unchanged when compared to the consistent use of {\texttt{PhenomXPHM}}. For our population, we calculated that our selection criteria reduced the overall cost of performing Bayesian inference on the population by $\sim 20\%$. While we only investigated the impact on the inferred mass and spin distributions for black holes in the Universe, we expect our selection criteria to reduce the computational cost and obtain indistinguishable results when \emph{e.g.} calculating the expansion rate of the Universe~\citep{LIGOScientific:2025jau} and identifying signatures of GW lensing~\citep{LIGOScientific:2025cwb}.

Although our current implementation involves obtaining the matched filter SNRs -- and hence choosing the model -- prior to sampling, it may be possible to include this calculation, and the selection criteria in the sampling itself, using the techniques developed in~\cite{Hoy:2024vpc}. Similarly, although our SNR threshold determines when precession and higher order multipole power is not needed for inference, it is possible that it could be extended to include eccentricity of the binaries orbit~\citep{Patterson:2024vbo,Singh:2025ojp}.

Throughout this work, we assumed a worse-case scenario population. For an astrophysically motivated population~\citep[based on the GW observations from the fourth GW transient catalog, GWTC-4,][]{LIGOScientific:2025slb}, we expect to obtain larger efficiencies since $> 90\%$ of observed binary black holes have lower spin magnitudes and more symmetric mass ratios where XAS is more likely to be selected. Even with the dominant usage of XAS, we expect to obtain consistent mass and spin distributions, see App.~\ref{sec:gwtclike_population} for details. From~\cite{Hoy:2024wkc}, 1 in 50 real GW signals show significant evidence for precession, and 1 in 70 events show significant evidence for higher order multipoles. As such, for a population of 500 real GW signals, we may expect to measure 10 events with significant evidence for precession and 8 with significant evidence for higher order multipoles. This implies that if our selection criteria was used, 482 events may be analysed with XAS without incurring any biases. This would reduce the overall cost of performing Bayesian inference on the population by up to $78\%$ compared to consistently using XPHM. In the limit that we consistently use XAS compared to XPHM, we would expect a reduction in total cost of 83\%. 

GW models are constantly improving in their accuracy to numerical relativity simulations; {\texttt{PhenomXPHM-SpinTaylor}}~\citep{Colleoni:2024knd}, {\texttt{PhenomXO4a}~\citep{Thompson:2023ase} and {\texttt{PhenomXPNR}}~\citep{Hamilton:2025xru} all show improvements beyond {\texttt{PhenomXPHM}}. However, as shown in Fig.~14 of~\cite{Hamilton:2025xru}, the waveform evaluation time typically increases with each successive improvement in the model. This means that by utilising our selection criteria, we expect the overall cost of performing Bayesian inference on the population to be reduced further with time.

It has been shown previously that combining marginal evidence through hierarchical inference can add up and provide a meaningful measurement, see e.g.~\cite{Stevenson:2017dlk}; for example combining many events with minimal evidence for a peak in the mass distribution can yield an informative measurement on the population. As such, there is a concern that analysing events with a model that excludes precession ignores this possibility. However, we note that a) we reconstruct an estimate for the spin tilt through our reweighting and conditioning procedure even when analysing a signal with a model that excludes precession, b) \cite{Wolfe:2025yxu} demonstrated that quiet events (marginal, low total SNR) do not contribute significantly on the population level (loud events contribute disproportionally), and their complete removal does not negatively impact the inferred mass and spin distribution of the population, and c) events with no evidence for precession and higher order multipoles will return exactly the prior, which when combined, will still give an uninformative measurement no matter the size of the catalog. As such, we believe our selection criteria, and the conclusions drawn in this work are robust to these effects.

Although the selection criteria described in this work is most suited for current and near-future detector networks~\citep{dcc:M1100296}, it could be used alongside next generation detectors\footnote{Unless novel methods are developed for performing Bayesian inference on GW data in $\mathcal{O}(\mathrm{sec})$, such as machine learning algorithms~\citep{Green:2020dnx,Green:2020hst,Dax:2021tsq} or others~\citep{Fairhurst:2023idl}}~\citep{Punturo:2010zza,Hild:2010id,ET:2019dnz,Reitze:2019iox,Babak:2021mhe,LISA:2024hlh} -- see \cite{Hoy:2024ovd} for a demonstration that a matched filtered $\rho_{\mathrm{HM}}$ can be extracted for signals observed with the Laser Interferometer Space Antenna~\citep[LISA,][]{LISA:2024hlh}. However, it may not provide significant speed ups since GW signals observed in next generation detectors are expected to be detected at large SNRs. This implies that most observations will show significant $\rho_{\mathrm{p}}$ and $\rho_{\mathrm{HM}}$ (although the exact number depends on the unknown mass, spin and redshift distributions of the expected sources). As a result, in most cases, highly accurate models will be required to avoid biases.

\section{Acknowledgments}
We are grateful to Mark Hannam for discussions and suggestions at the start of this project, Konstantin Leyde for conversations about hierarchical Bayesian inference and Matthew Mould for helpful comments during the LIGO-Virgo-KAGRA internal review. We thank Stephen Fairhurst, Ian Harry, Laura Nuttall and Mukesh Singh for valuable feedback on this manuscript. We additionally thank Stephen Fairhurst and Mukesh Singh for their continued development of {\texttt{simple-pe}} (alongside C.H). We thank the University of Portsmouth for support through the Dennis Sciama Fellowship, and we are grateful for computational resources provided by the SCIAMA high performance computing cluster which is supported by the Institute of Cosmology and Gravitation (ICG) and the University of Portsmouth. For the purpose of open access, the author(s) has applied a Creative Commons Attribution (CC BY) licence to any Author Accepted Manuscript version arising.

Plots were prepared with Matplotlib~\citep{2007CSE.....9...90H}, {\texttt{pesummary}}~\citep{Hoy:2020vys} and {\texttt{gwpopulation}}~\citep{2019PhRvD.100d3030T}. Single-event parameter estimation was performed with {\texttt{bilby}}~\citep{Ashton:2018jfp,Romero-Shaw:2020owr}, and hierarchical parameter estimation was performed with {\texttt{gwpopulation}}~\citep{2019PhRvD.100d3030T}; both making use of the {\texttt{dynesty}} nesting sampling package~\citep{Speagle:2019ivv}. {\texttt{simple-pe}}~\citep{Fairhurst:2023idl} and {\texttt{pesummary}}~\citep{Hoy:2020vys} were used to extract the matched filter $\rho_{\mathrm{p}}$ and $\rho_{\mathrm{HM}}$ from the GW strain data. {\texttt{NumPy}}~\citep{harris2020array} and {\texttt{SciPy}}~\citep{2020SciPy-NMeth} were also used in our analyses.

\section{Data Availability}

Code to evaluate our selection criteria, as well as all posterior samples obtained in this work (both single-event and hierarchical analyses) are made publicly available on \href{https://github.com/icg-gravwaves/reconsidering_PHM_models_for_inference}{GitHub}.

\bibliographystyle{mnras}
\bibliography{main}

\appendix
\section{Methods}

\subsection{Single event Bayesian inference} \label{sec:bayesian_inference}

Bayesian inference is the process of inferring the properties $\theta$ of a model $\mathfrak{M}$ given some observed data $d$. The probability of \emph{e.g.} the binary black hole having parameters $\theta$ is obtained through Bayes theorem,

\begin{equation}
    P(\theta | d, \mathfrak{M}) = \frac{\mathcal{L}(d | \theta, \mathfrak{M})\,\Pi(\theta | \mathfrak{M})}{\mathcal{Z}},
\end{equation}
where $\mathcal{L}(d | \theta, \mathfrak{M})$ is the probability of the data given the parameters $\theta$ and model $\mathfrak{M}$, otherwise known as the likelihood, $\Pi(\theta | \mathfrak{M})$ is the probability of the parameters $\theta$ given the model $\mathfrak{M}$, otherwise known as the prior and $\mathcal{Z} = \int(\mathcal{L}(d | \theta, \mathfrak{M})\,\Pi(\theta | \mathfrak{M}))d\theta$ is the evidence or marginal likelihood. When Bayesian inference is performed with multiple models, the evidence can be used to quantify the support for one model over the other through the Bayes factor,

\begin{equation}
    \mathcal{B} = \frac{\mathcal{Z}_{1}}{\mathcal{Z}_{2}},
\end{equation}
where $\mathcal{Z}_{1}$ is the evidence for $\mathfrak{M}_{1}$ and $\mathcal{Z}_{2}$ is the evidence for $\mathfrak{M}_{2}$. An alternative technique is to sample over the models as part of the inference~\citep{Hoy:2022tst}. This has the advantage that model accuracy can then be incorporated~\citep{Hoy:2024vpc}.

It is often not possible to analytically calculate the posterior distribution. Instead, stochastically sampling techniques, such as nested sampling~\citep{Skilling2004,Skilling:2006gxv} or Markov Chain Monte-Carlo~\citep{metropolis1949monte}, are often employed to obtain \emph{samples} from the unknown posterior distribution; although alternative methods have been proposed in the literature~\citep{Pankow:2015cra,Lange:2018pyp,Delaunoy:2020zcu,Green:2020hst,Chua:2019wwt,Green:2020dnx,Dax:2021tsq,Gabbard:2019rde,Tiwari:2023mzf,Fairhurst:2023idl,Williams:2025szm}. Numerous packages are available for performing Bayesian inference on GW signals~\citep{Veitch:2014wba,Ashton:2018jfp,Romero-Shaw:2020owr,Biwer:2018osg}. In this work, we use {\texttt{bilby}}~\citep{Ashton:2018jfp,Romero-Shaw:2020owr}, and employ the {\texttt{dynesty}}~\citep{Speagle:2019ivv} nested sampler.

When analysing simulated GW signals $h$ injected into idealised Gaussian noise $n$ -- such that $d = h + n$ -- we use wide and agnostic priors that follow the conventions adopted by the LVK. Specifically in this work, we sample uniformly in component masses between a chirp mass $2\,M_{\odot} < \mathcal{M} < 200\, M_{\odot}$ and mass ratio $0.05 < q < 1$. The luminosity distance is also assumed to be uniform in volume between $10\,\mathrm{Mpc} < d_{\mathrm{L}} < 10000\,\mathrm{Mpc}$. All other parameters are assumed to be agnostic within their regions of validity. Given the wide prior volume, we use 2000 live points to ensure robust results. Although numerous methods have been developed to accelerate likelihood evaluations, we use the full Whittle likelihood in this work.

\subsection{Population inference} \label{sec:pop_inference}

Once a catalog of observations has been made $\mathcal{D}$, Hierarchical Bayesian inference can be used to infer the underlying properties of the population $\lambda$. The probability of \emph{e.g.} the catalog of binary black holes following a mass and spin distribution with hyperparameters $\lambda$ is similarly obtained through Bayes theorem,

\begin{equation}
    P(\lambda | \mathcal{D}) = \frac{\mathcal{L}(\mathcal{D} | \lambda)\, \Pi(\lambda)}{p(\mathcal{D})},
\end{equation}
where $\mathcal{L}(\mathcal{D} | \lambda)$ is the hierarchical likelihood, $\Pi(\lambda)$ is the hyperparameter prior, and $p(\mathcal{D})$ is evidence. Assuming all GW signals are of astrophysical origin and all noise realizations are independent, the hierarchical likelihood is~\citep{Mandel:2018mve,Vitale:2020aaz,LIGOScientific:2025yae},

\begin{equation} \label{eq:pop_inf}
    \mathcal{L}(\mathcal{D} | \lambda) = \prod_{i = 0}^{N} \frac{\int{P(d | \theta)}\, P(\theta | \lambda) d\theta}{\int{p_{\mathrm{det}}(\theta) P(\theta | \lambda)}}
\end{equation}
where $N$ is the number of observed events in the catalog, $d = \mathcal{D}_{i}$ and corresponds to the $i\mathrm{th}$ event in the catalog, $P(\theta | \lambda)$ specifies the distribution of binary black hole parameters $\theta$ given the mass and spin model with parameters $\lambda$, and $p_{\mathrm{det}}(\theta)$ is the detection probability of an event with parameters $\theta$. It is crucial to include the detection probability in the denominator as current template-based search pipelines adopted by the LVK~\citep{Cannon:2011vi, Privitera:2013xza,Messick:2016aqy,Hanna:2019ezx,Sachdev:2019vvd,Usman:2015kfa,Nitz:2017svb,Nitz:2018rgo,Adams:2015ulm,Chu:2020pjv,Liu:2012vw,Guo:2018tzs} are more likely to observe GW signals from \emph{e.g.} high mass binaries than low mass binaries due to the increase in sensitive volume. The integrals in Eq.~\ref{eq:pop_inf} are typically approximated through Monte Carlo sums~\citep{Tiwari:2017ndi}.

Often the detection probability is determined by injecting $\mathcal{O}(10^{8})$ simulated GW signals into real detector noise, and identifying the fraction of signals identified from the (currently adopted) search pipelines~\citep{LIGOScientific:2025pvj}. In this work, we simplify the process by assuming all signals with total SNR greater than 12 are detected~\citep{Essick:2023upv,Agarwal:2024hld,Hoy:2025ule}. While the detection probability depends on the waveform model (used when simulating the GW signals), we expect any differences to be small. We therefore neglect this systematic error. 

When performing Hierarchical Bayesian inference in this work, we assume a {\textsc{PowerLaw+Peak}} distribution~\citep{LIGOScientific:2020kqk} for the primary component mass and a powerlaw for the mass ratio. The spin magnitudes are assumed to be independently and identically drawn from a Beta distribution~\citep{Wysocki:2018mpo}, and a mixture model is used for the spin tilt~\citep{Talbot:2018cva}. Similar to single-event Bayesian inference, we use stochastic sampling to obtain posterior samples from the unknown hyperposterior. We use {\texttt{gwpopulation}}~\citep{2019PhRvD.100d3030T}, and employ the {\texttt{dynesty}}~\citep{Speagle:2019ivv} nested sampler with 1000 live points. We assume wide and agnostic priors for all hyper parameters $\lambda$. Specifically, we assume that the powerlaw for the primary mass is uniform between $-4 < \alpha < 12$ and the powerlaw for the mass ratio is uniform between $-5 < \beta_{q} < 12$. The minimum primary mass is assumed to be uniform between $2\,M_{\odot{}} < m_{\mathrm{min}} < 10\,M_{\odot}$, while the maximum is allowed to vary uniformly between $30\, M_{\odot} < m_{\mathrm{max}} < 100\, M_{\odot}$. For the spin magnitude distribution, we assume both parameters of the Beta distribution are uniform between $0 < \alpha_{\chi} < 10$ and $0 < \beta_{\chi} < 10$. 

\section{Alternative SNR thresholds} \label{sec:different_thresholds}

\begin{table}
    \begin{center}
    \begin{tabular}{| c | c |}
    \hline
    \hline
    \multirow{2}{*}{Threshold ($\rho_{\mathrm{thres}}$)} & Reduction in \\
    & computational cost [\%] \\
    \hline
    $0$ & - \\
    $0.5$ & 2 \\
    $1.0$ &  6 \\
    $1.5$ & 20 \\
    $2.0$ & 25 \\
    $2.5$ &  40 \\
    $3.0$ &  45 \\
    \hline
    \hline
    \end{tabular}
    \end{center}
    \caption{Table showing the reduction in computational cost of performing Bayesian inference for a worst-case scenario population of 90 events for different SNR thresholds, $\rho_{\mathrm{thres}}$. We compare to the computational cost when consistently using XPHM for all events, $\rho_{\mathrm{thres}} = 0$. For $\rho_{\mathrm{thres}} = 0$, the total computational cost of analysing our population was 50 CPU years.}
    \label{tab:runtimes}
\end{table}

\begin{figure*}
    \centering
    \includegraphics[width=0.94\textwidth]{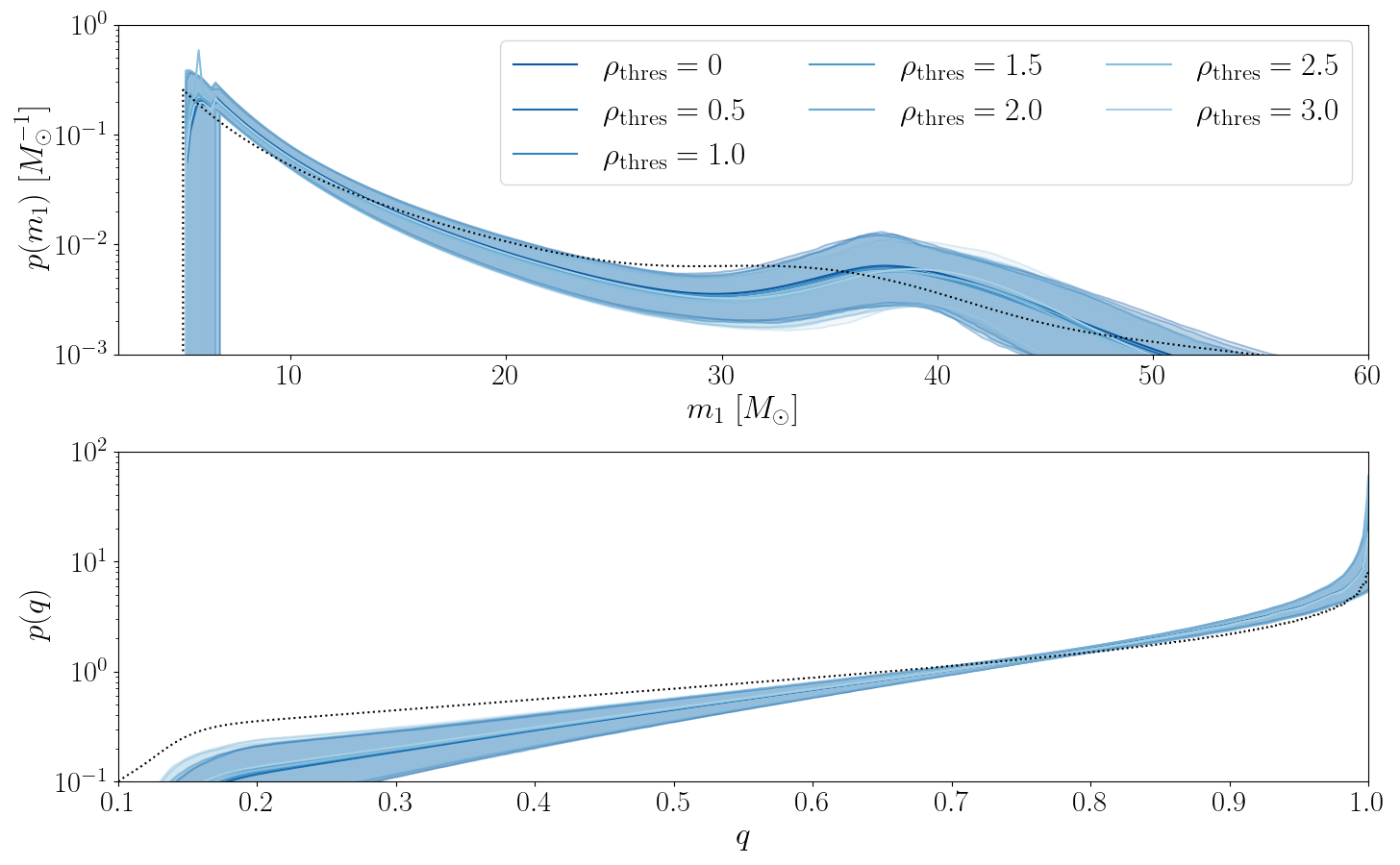}
    \includegraphics[width=0.94\textwidth]{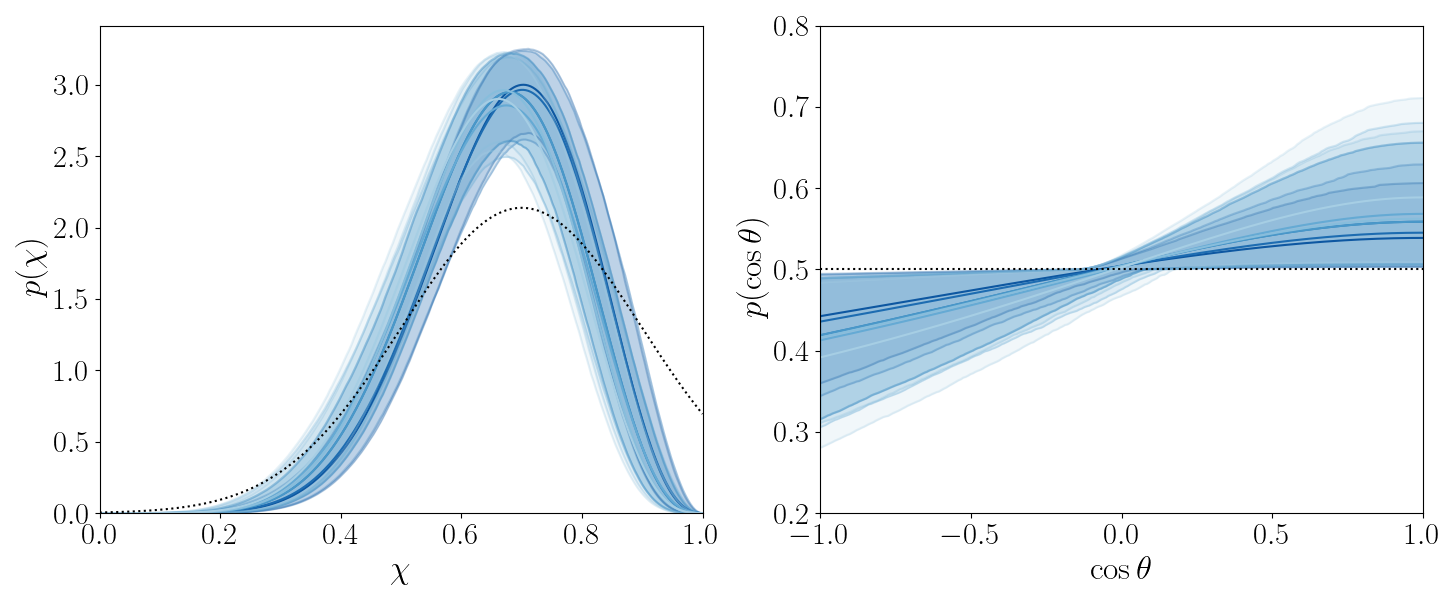}
    \caption{Plot similar to Fig.~\ref{fig:pop_inf} except we now show the inferred mass and spin distribution for different SNR thresholds $\rho_{\mathrm{thres}}$. In this work, we recommend using $\rho_{\mathrm{thres}} = 1.5$ for robust results. $\rho_{\mathrm{thres}} = 0$ corresponds to XPHM in Fig.~\ref{fig:pop_inf}.}
    \label{fig:pop_inf_multiple_snrs}
\end{figure*}

In the main text, we highlighted the inferred mass and spin distribution for two SNR thresholds, $\rho_{\mathrm{thres}} = 1.5$ and $\rho_{\mathrm{thres}} = 2.0$. Here, we show the results for SNR thresholds $\rho_{\mathrm{thres}} \in [0, 0.5, 1.0, 1.5, 2.0, 2.5, 3.0]$.

In Fig.~\ref{fig:pop_inf_multiple_snrs} we see that the inferred mass distribution remains comparable for different SNR thresholds. Focusing on the mass ratio distribution, we see that as the threshold increases, there is marginally more support for asymmetric binaries; $\beta_{q}$ has larger support for negative values as $\rho_{\mathrm{thres}}$ increases. This implies that even for binaries with some evidence for precession and higher order multipoles, XAS can be used (alongside other models for candidates with significant evidence for precession and higher order multipoles), and is sufficient for capturing the mass distribution. This is consistent with the conclusions found in~\cite{Singh:2023aqh}. As the threshold increases further, such that XAS is being used for all GW candidates, biases begin to be observed in the mass distribution, particularly in the inferred powerlaw of the primary mass. This is consistent with the conclusions found in~\cite{Hoy:2025ule}.

\setcounter{section}{3}
\setcounter{figure}{0}
\begin{figure*}
    \centering
    \includegraphics[width=0.92\textwidth]{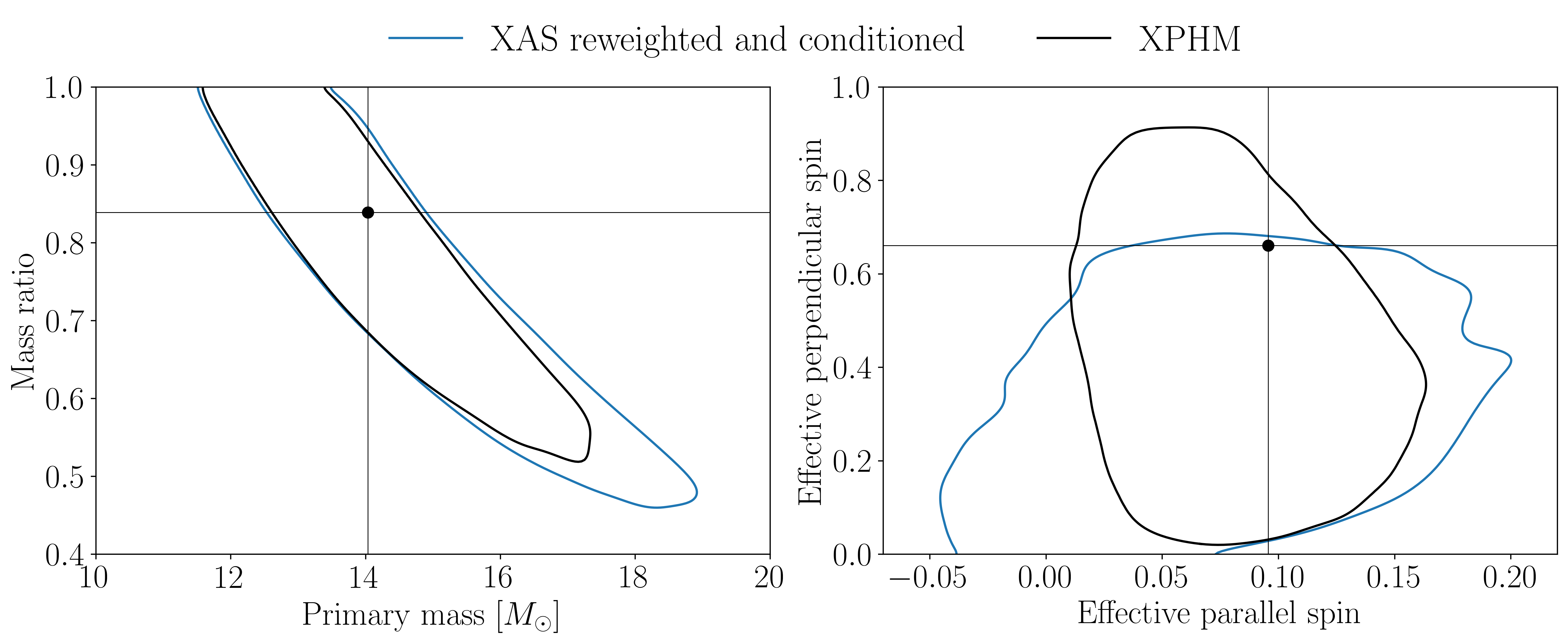}
    \caption{Plot showing the inferred \emph{Left}: primary mass and mass ratio and \emph{Right}: effective parallel spin $\chi_{\mathrm{eff}}$ and effective perpendicular spin $\chi_{\mathrm{p}}$ when performing Bayesian inference on a misidentified GW signal; a signal that we identify as having no evidence for spin-precession despite there being  strong observable features. This injection corresponds to a false negative in Fig.~\ref{fig:snr_accuracy}. The GW signal was produced with XPHM into idealised Gaussian noise. The black cross hairs show the true values and the contours show the inferred 90\% credible interval. The reweighting and conditioning technique applied to XAS is described in Sec.~\ref{sec:verification} and Fig.~\ref{fig:rweighting}.}
    \label{fig:misidentified_signal}
\end{figure*}
\setcounter{section}{2}
\setcounter{figure}{1}

The same is not true for the inferred spin distribution. We see that as $\rho_{\mathrm{thres}}$ increases, biases are observed in the spin magnitude and spin tilt distributions, with the spin magnitude marginally underestimated, and the spin tilts favouring more aligned configurations. This is expected from the reweighting and conditioning technique we adopted to \emph{predict} the spin tilt distribution for XAS, see Fig.~\ref{fig:rweighting}. Since we are only conditioning on the observed $\chi_{\mathrm{eff}}$, we are neglecting any measurable power from the in-plane spins and any contribution from the mass ratio~\citep{Baird:2012cu}.

In Tab.~\ref{tab:runtimes} we provide a comparison of the runtimes for different SNR thresholds. We see that as the threshold increases, there is a greater reduction in computational cost compared to consistently using XPHM for all analyses. This is expected since a greater fraction of events will be analysed with computationally cheaper models.

\section{A misidentified signal} \label{sec:misidentified_signal}

In Sec.~\ref{sec:selection_criteria}, we highlighted that {\texttt{simple-pe}} can be used to directly extract $\rho_{\mathrm{p}}$ and $\rho_{\mathrm{HM}}$ from the GW strain data through matched filtering. We then compared the true and matched filter SNRs and highlighted that in 6\% of cases, we obtain false negatives. This means that our selection criteria may have mis-selected XAS compared to XPHM, XP or XHM. Here, we show the impact of misidentifying the significance of the signal on the inferred parameter estimates. We consider an injection that has significant evidence for spin-precession, $\rho_{\mathrm{p}} \sim 3.5$, yet we obtain a matched filter SNR of $\sim 1.5$ through {\texttt{simple-pe}}. This injection had component masses $m_{1} = 14\, M_{\odot}, m_{2} = 13\, M_{\odot}$, spin magnitudes $\chi_{1} = 0.6, \chi_{2} = 0.8$ and observed at a network SNR of 17 in LIGO-Hanford and Virgo. The effective parallel spin was $\chi_{\mathrm{eff}} = 0.1$ and the effective perpendicular spin was $\chi_{\mathrm{p}} = 0.7$.

Fig.~\ref{fig:misidentified_signal} shows the inferred posterior distribution for XPHM and XAS, after reweighting to a uniform spin magnitude prior and conditioning based on the observed $\chi_{\mathrm{eff}}$, see Sec.~\ref{sec:verification} and Fig.~\ref{fig:rweighting} for details. We see that XAS and XPHM both recover the primary mass and mass ratio well. XPHM obtains a more precise distribution, as expected. We see a larger difference for the inferred effective spins, but importantly both XAS and XPHM recover the true value within the 90\% credible interval. We see that XAS obtains a wider distribution for $\chi_{\mathrm{eff}}$ with less support for larger effective perpendicular spins. This is expected from the reweighting and conditioning procedure, as we ignore any measurable evidence of precession.

The inferred 90\% credible interval scales with the total SNR: as the SNR increases, the inferred 90\% credible interval narrows for all models. If this injection had SNR $> 20$ it is likely that XAS would recover a biased distribution for the spins. This begs the question of whether or not we should include the network SNR when determining the selection threshold. For example, we could always use XPHM if the total SNR is greater than \emph{e.g.} 20, even if the matched filter $\rho_{\mathrm{p}} < \rho_{\mathrm{thres}}$ and $\rho_{\mathrm{HM}} < \rho_{\mathrm{thres}}$ to avoid this issue. This would also alleviate additional concerns from the improved measurement from XPHM due to no evidence for precession and higher order multipoles, as discussed in Sec.~\ref{sec:models}. However, we highlight that even without this additional constraint, we obtain comparable mass and spin distributions for the underlying properties of black holes for a worst-case scenario population. This implies that this additional constraint, which would reduce the computational efficiency of our approach, may not be necessary for population level measurements.

\section{A GWTC-like population} \label{sec:gwtclike_population}

In the main text, we considered a worst-case scenario population containing highly spinning binary black holes with a preference for asymmetric component masses. Here, we consider an astrophysically motivated population of binary black holes. This population represents the signals observed by the LVK: a `GWTC-like' population. Since this population contains fewer events with observable spin-precession and higher order multipole power~\citep{Hoy:2024wkc}, we expect to obtain consistent mass and spin distribution when imposing our recommended SNR threshold of $\rho_{\mathrm{thres}} = 1.5$. This is further supported by the conclusions in~\cite{Singh:2023aqh,Hoy:2025ule}; Fig.~5 in \cite{Singh:2023aqh} demonstrated that neglecting higher order multipoles will not cause a significant bias in the population inference for a catalog of 750 simulated signals, and Figs.~3 and~9 in~\cite{Hoy:2025ule} demonstrated that for an astrophysically motivated population of 100 signals, neglecting precession will not cause a significant bias in the population inference.

We obtain our simulated population by assuming that the mass ratio follows a power law $q^{\beta_{q}}$ with index $\beta_{q} = 1.1$, and a uniform distribution for the spin magnitudes. We also assume that the primary mass follows a truncated power law with index $\alpha=-2.3$, between a minimum, $m_{\mathrm{min}} = 5\, M_{\odot}$,and maximum $m_{\mathrm{max}} = 80\, M_{\odot}$ mass. We assume that there is a Gaussian component in the mass distribution at $m_{1} = 34\, M_{\odot}$, with relative weight $\lambda = 0.038$ and standard deviation $\sigma_{\rm g} = 5~M_{\odot}$.

\begin{figure}
    \includegraphics[width=0.48\textwidth]{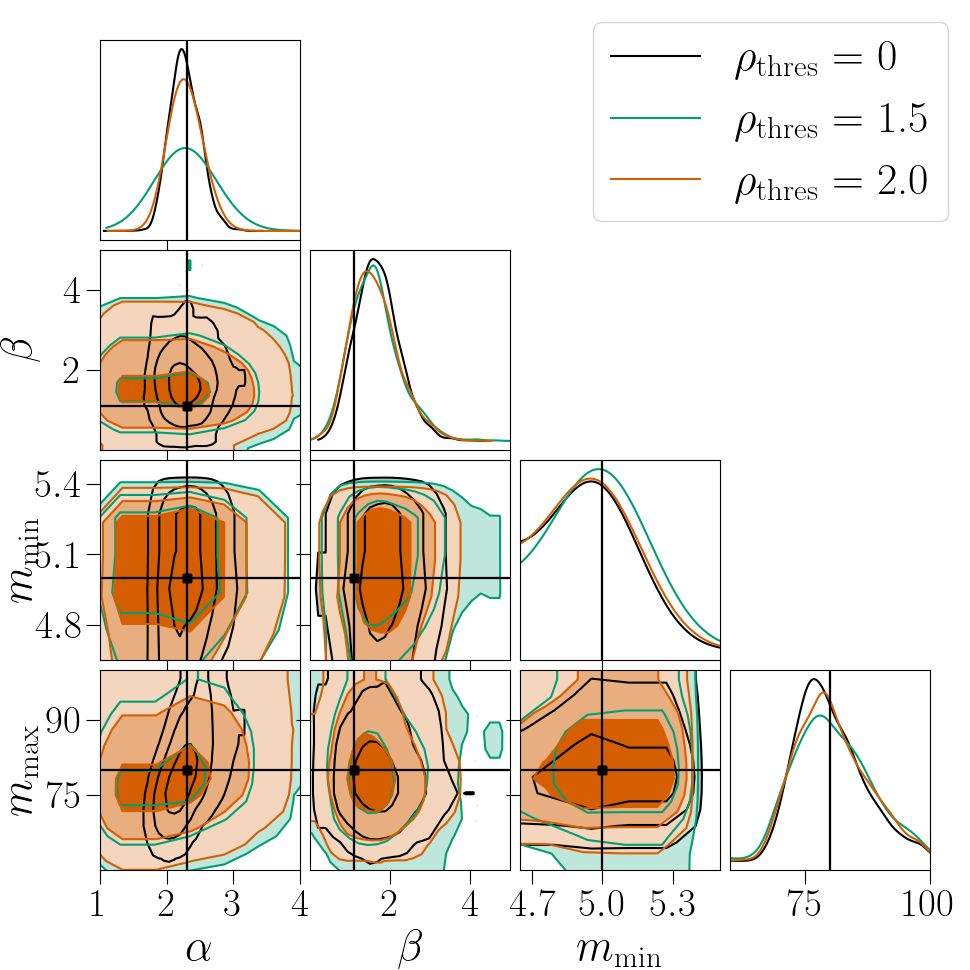}
    \caption{Corner plot showing a subset of the inferred hyper parameters controlling the mass distribution of black holes. We consider a GWTC-like population with a uniform distribution of spins, see App.~\ref{sec:gwtclike_population} for details. The black cross hairs show the true values.}
    \label{fig:mass_corner_GWTC_like}
\end{figure}

As expected, we obtain conclusions similar to those in the main text: the mass distribution remains comparable to XPHM for the SNR thresholds considered. However, we see marginal differences in the spin distribution. Namely, a preference for more aligned spin binaries. In Fig.~\ref{fig:mass_corner_GWTC_like} we show a corner plot comparing the hyper posteriors for the mass distribution against their true values.

\bsp 
\label{lastpage}
\end{document}